\documentclass[preprint,review,12pt]{elsarticle}
\usepackage{amssymb}
\usepackage{graphicx}
\journal{Physics Letters B}

\def\aA{$\alpha$-nucleus\ }
\def\aap{($\alpha,\alpha')^{12}$C\ }
\def\ppp{($p,p')^{12}$C\ }
\def\ccp{($^{12}$C,$^{12}$C$^*)^{12}$C\ }
\def\AA{nucleus-nucleus\ }
\def\aC{$\alpha+^{12}$C\ }

\def\ac{$\alpha+^{12}$C }
\def\aPb{$\alpha+^{208}$Pb\ }

\begin{document}
\begin{frontmatter}
\title{Hindrance of the excitation of the Hoyle state and the ghost of the
 $2^+_2$ state in $^{12}$C}
\author{Dao T. Khoa$^1$, Do Cong Cuong$^1$, Yoshiko Kanada-En'yo $^2$}
\address{$^{1}$Institute for Nuclear Science {\rm \&} Technique, VAEC \\
179 Hoang Quoc Viet Rd., Nghia Do, Hanoi, Vietnam. \\
$^2$ Department of Physics, Kyoto University, Kyoto 606-8502, Japan.}
\begin{abstract}
While the Hoyle state (the isoscalar $0^+_2$ excitation at 7.65 MeV in $^{12}$C)
has been observed in almost all the electron and $\alpha$ inelastic scattering experiments,
the second $2^+$ excited state of $^{12}$C at $E_{\rm x}\approx 10$ MeV, believed to be
an excitation of the Hoyle state, has not been clearly observed in these measurements
excepting the high-precision  \aap experiments at $E_\alpha=240$ and 386 MeV.
Given the (spin and isospin zero) $\alpha$-particle as a good probe for the nuclear
isoscalar excitations, it remains a puzzle why the peak of the $2^+_2$ state could
not be clearly identified in the measured \aap spectra.
To investigate this effect, we have performed a microscopic folding model analysis of the
\ac scattering data at 240 and 386 MeV in both the Distorted Wave Born Approximation
(DWBA) and  coupled-channel (CC) formalism, using the nuclear transition densities
given by the antisymmetrized molecular dynamics (AMD) approach and a complex
CDM3Y6 density dependent interaction.
Although AMD predicts a very weak transition strength for the direct $(0^+_1\to 2^+_2)$
excitation, our detailed analysis has shown evidence that a weak \emph{ghost} of the $2^+_2$
state could be identified in the 240 MeV \aap data for the $0^+_3$ state at 10.3 MeV,
when the CC effects by the indirect excitation of the $2^+_2$ state are taken into account.
Based on the same AMD structure input and preliminary \aap data at 386 MeV, we have
estimated relative contributions from the $2^+_2$ and $0^+_3$ states to  the excitation
of $^{12}$C at $E_{\rm x}\approx 10$ MeV as well as possible contamination by $3^-_1$
state.
\end{abstract}
\end{frontmatter}

The excited states of $^{12}$C lying around the $\alpha$-decay threshold
have become a research subject of wide interest recently \cite{Fre07,Fun09} because of the
dominant $\alpha$-cluster structure established in some cases, such as the isoscalar 0$^+_2$ state
at 7.65 MeV in $^{12}$C (known as the Hoyle state that has a vital role in the stellar synthesis
of Carbon). Although the three $\alpha$-cluster structure of the Hoyle state has been shown more
than 30 years ago in the microscopic Resonating Group Method (RGM) calculations
\cite{Ueg77,Kam81,Pic97}, an interesting $\alpha$-condensate scenario \cite{Fun09}
for this state has been established just recently \cite{Toh01,Fun03}, where three $\alpha$
clusters were shown to condense into the lowest $S$ state of their potential. A more complicated
structure of the Hoyle state was found in the Fermionic Molecular Dynamics (FMD) calculation
\cite{Che07} where the condensate wave function is mixed also with the molecular $^8$Be$+\alpha$
configuration, but the condensate component still exhausts about 70\% of the total wave function.
Given such a strong condensate of the three $\alpha$ clusters, a question arises naturally about
the isoscalar (IS) excitation of the Hoyle state. Namely, if it is a condensate $S$ state then
the next level in the potential containing three $\alpha$-particles should be a $D$ state
formed by promoting an $\alpha$-particle from the $S$ to $D$ level. Such an excited state
has been first predicted by Funaki {\it el al.} \cite{Fun05} and it must be a 2$^+$ state at the
excitation energy of around 10 MeV, with a pronounced $^8$Be+$\alpha$ structure \cite{Fre07}.
This same $2^+_2$ state has been predicted also by the three-body calculation \cite{Kato} or the
antisymmetrized molecular dynamics (AMD) approach \cite{Enyo07}, as the second $2^+$ state
of $^{12}$C lying about 2 MeV above the  $\alpha$-decay threshold. The experimental
observation of the $2^+_2$ state of $^{12}$C would be very important for a deeper
understanding of the Hoyle state, e.g., the measured excitation energy would allow us
to determine the moment of inertia and deformation of $^{12}$C being in the Hoyle state.
The first experimental hint for the $2^+_2$ state has been found by the Texas A\&M
University group in the isoscalar $E2$ strength distribution of $^{12}$C in the energy range
$10 \lesssim E_{\rm x}\lesssim 30$ MeV \cite{John03}. However, this $2^+$ peak is located
at $E_{\rm x}\approx 11.46\pm 0.20$ MeV which is somewhat high compared to the predicted
value around 10 MeV. A more convincing experimental measurement of the $2^+_2$ state has been
performed by Itoh {\it et al.} in the 386 MeV inelastic \aC scattering spectrum  \cite{Itoh04,Itoh08},
based on a multipole decomposition analysis (MDA) of the measured \aap angular distribution.
Given a prominent 3-$\alpha$ cluster structure predicted for this state, several experimental
efforts \cite{Fre07a,Diget07} have also been made by Freer {\it et al.} to search for the
$2^+_2$ peak in the 3-$\alpha$ decay spectrum of $^{12}$C in the excitation energy range
of $9\lesssim E_{\rm x}\lesssim 11$ MeV but no positive identification has been done.
Recently, Freer and collaborators have performed the \ppp experiment at the beam energy
of 66 MeV \cite{Fre09} as well as the \ccp experiment at 101.5 MeV \cite{Bri10}.
While some enhancement above background has been deduced from the \ppp spectrum that indicates
a possible $2^+_2$ peak at $9.6\pm 0.1$ MeV \cite{Fre09},  no conclusive evidence was found
in the latter experiment excepting some estimate made for the upper limits in the excitation strength
of the $2^+_2$ state  \cite{Bri10}.  The present work is our attempt to shed some light into this
puzzled situation by a detailed folding model analysis of inelastic \aC scattering data at
240 MeV \cite{John03} and 386 MeV \cite{Itoh04,Itoh08}.

Because the spin- and isospin zero $\alpha$-particle is a very good projectile to excite
the nuclear IS states, the 3-$\alpha$ RGM wave function obtained by Kamimura \cite{Kam81}
has been used earlier in the folding model analysis \cite{Kho08} of the inelastic \ac scattering to
probe the $E0$ transition strength of the Hoyle state. This approach has been extended to study
also other IS excitations of $^{12}$C like 2$^+$ (4.44 MeV), 3$^-$ (9.64 MeV), 0$^+$ (10.3 MeV)
and 1$^-$ (10.84 MeV) states \cite{Kho08k}, using the same RGM wave functions.
The technical details of this folding approach for elastic and inelastic \AA scattering can be found
in Ref.~\cite{Kho00}. The key quantity in our folding model analysis is the \aA form factor (FF)
that contains all the information about the \aA inelastic scattering as well as structure of the
nuclear state under study. Therefore, it is vital to evaluate the FF using a good choice for the
effective nucleon-nucleon (NN) interaction and realistic wave functions for the $\alpha$-particle
and target nucleus, respectively. In the present work, we apply our folding model approach to study
the possible excitation of the $2^+_2$ state of  $^{12}$C using the microscopic nuclear transition
densities given by the AMD calculation \cite{Enyo07} and  the (complex)  density-dependent
CDM3Y6 interaction, whose parameters have been fine tuned recently \cite{Kho10} for the \aA
scattering at the same incident energies of 240 and 386 MeV.

The AMD approach has been proven to be quite reliable in describing the structure of low-lying
excited states in light nuclei, where both the cluster and shell-model like states are consistently
reproduced \cite{Enyo07,Enyo95}. In the present work, the structure of IS excited states of $^{12}$C
is generated within the AMD approach using the method of variation after the spin-parity
projection (VAP).  The main structure properties of these states are summarized in Table~\ref{t1}.
While the AMD prediction for the shell-model like $2^+_1$ state is quite satisfactory in both
the excitation energy and $E2$ transition strength, the predicted excitation energies for higher
lying states are slightly larger than the experimental values. However, such a difference in the excitation
energies does not affect significantly the calculated inelastic \aC scattering cross section because
it can lead only to a very small change in the kinetic energy of emitted $\alpha$-particle and, thus,
can be neglected.  Most vital are the strength and shape of the nuclear transition density used
to evaluate the inelastic FF that can affect directly the inelastic scattering cross section calculated
in the Distorted Wave Born Approximation (DWBA) or coupled-channel (CC) formalism.
The details of the AMD calculation for the IS excited states of $^{12}$C are given in
Ref.~\cite{Enyo07}. In the present work, the AMD nuclear transition densities enter the
folding calculation in the same convention as in Refs.~\cite{Kho00,Kho10} so that the
isoscalar transition strength for a $2^\lambda$-pole nuclear transition
$|J_i\rangle\to |J_f\rangle$ is described by the reduced nuclear transition rate
$B({\rm IS}\lambda;J_i\to J_f)=|M({\rm IS}\lambda;J_i\to J_f)|^2$, where the
$2^\lambda$-pole transition moment is determined from the corresponding nuclear transition
density as
\begin{eqnarray}
M({\rm IS}\lambda;J_i\to J_f)&=&\int dr\ r^{\lambda+2}
 \rho_{J_f,J_i}^{(\lambda)}(r) \ \ \  {\rm if}\ \ \  \lambda\geqslant 2, \label{dens} \\
M({\rm IS}0;J_i\to J_f)&=&\int dr\ r^{4} \rho_{J_f,J_i}^{(\lambda=0)}(r),
 \label{dens0} \\
M({\rm IS}1;J_i\to J_f)& =& \int dr\left(r^{3}-
 {5\over 3}\langle r^2\rangle r\right)r^2 \rho_{J_f,J_i}^{(\lambda=1)}(r).  \label{dens1}
\end{eqnarray}
Note that the IS dipole transition moment is evaluated based on higher-order corrections to
the dipole operator, with spurious center-of-mass (c.m.) oscillation subtracted \cite{giai}.
The reduced electric transition rate is evaluated as
$B(E\lambda;J_i\to J_f)=|M(E\lambda;J_i\to J_f)|^2$, where $M(E\lambda)$ is determined
in the same way as $M({\rm  IS}\lambda)$ but using the proton part of the nuclear transition
density only. We will discuss hereafter the transition strength in terms of $B(E\lambda)$ only
because this is the quantity that can be compared with the experimental data whenever possible.

The excitation energies and $E\lambda$ transition strengths of the IS states considered
in the present work are given in Table~\ref{t1}. One can see that the calculated excitation
energies and $E\lambda$ transitions from the ground state $0^+_1$ to  the $2^+_1$,  $0^+_2$
and $3^-_1$ states agree reasonably with the experimental values. As illustrated in
Figs.~\ref{Elsig2} and \ref{Insig3} below, the AMD nuclear transition densities also give good
description of the corresponding inelastic \aap cross sections. In difference from the shell-model
like structure of the $2^+_1$ state, the $2^+_2$ state has a well established cluster structure
(see Fig.~5 of Ref.~\cite{Enyo07}), with a more extended and dilute mass distribution that
corresponds to the mass radius $R_{\rm m}\approx 3.99$ fm which is even larger than
that of the Hoyle state. The more striking are the predicted electric transition rates for the
$E2$ transitions  from the Hoyle state to the $2^+_2$ state and from the $2^+_2$ state
to the $4^+_2$ state: $B(E2;0^+_2\rightarrow 2^+_2)\approx 511\  e^2$fm$^4$ and
$B(E2;2^+_2\rightarrow 4^+_2)\approx 1071\  e^2$fm$^4$ that are much stronger than
those of the $E2$ transitions between the members of the ground-state rotational band:
$B(E2;0^+_1\rightarrow 2^+_1)\approx 42.5\  e^2$fm$^4$
and $B(E2;2^+_1\rightarrow 4^+_1)\approx 28.5\  e^2$fm$^4$. As a result, the predicted
$B(E2;0^+_2\rightarrow 2^+_2)$ and $B(E2;2^+_2\rightarrow 4^+_2)$ transition rates
strongly suggest that the $2^+_2$ and $4^+_2$ states should be the members of the
excited rotational band built upon the Hoyle state. The $B(E2;0^+_2\rightarrow 2^+_2)$
values predicted by the RGM \cite{Kam81} and  FMD calculations \cite{Neff} are even
larger than that given by the AMD calculation. Given a very weak direct excitation of
the $2^+_2$ state from the ground state, $B(E2;0^+_1\rightarrow 2^+_2)\approx
2\  e^2$fm$^4$ predicted by the AMD calculation, we can draw a conclusion that the
$2^+_2$ state should be an IS quadrupole excitation of the Hoyle state \cite{Fre07}.
It should be noted that if we take the measured $E2$ strength of the $2^+_2$
peak at 11.46 MeV in the 240 MeV \aap spectrum, which exhausts $2.15\pm 0.30\%$
of the $E2$ energy weighted sum rule (EWSR) \cite{John03}, then we obtain
$B(E2;0^+_1\rightarrow 2^+_2)_{\rm exp}\approx 2.5\pm 0.5\  e^2$fm$^4$
based on the standard collective model treatment of the MDA \cite{Kho10}.
This value agrees surprisingly well with that predicted by the AMD calculation
and it is, therefore, not excluded that the observed  $2^+$ peak at $E_{\rm x}\approx
11.46$ MeV in the 240 MeV \aap spectrum corresponds to the $2^+_2$ state,
although the excitation energy is about 1 MeV above the value predicted by the AMD.
The width of this state has been determined from the 240 MeV spectrum to be
$\Gamma_{\rm c.m.}\approx 430\pm 100$ keV \cite{John03}, which is somewhat
smaller than that ($\sim 600$ keV) suggested by Freer {\it et al.} \cite{Fre09}.
A closer look indicates that the $2^+$ peak at 11.46 MeV in the 240 MeV
spectrum might well be the adopted ($2^+$) level of $^{12}$C \cite{Selo85} at
$E_{\rm x}\approx 11.16\pm 0.05$ MeV having a width of $550\pm 100$ keV,
observed in the ($^3$He,$d$) stripping reaction at $E_{\rm lab}= 44$ MeV
\cite{Rey71}.  It should be noted, however, that this state has only been
seen once in the $^{11}$B($^3$He,$d$) reaction, and not in other studies. Therefore,
it is not excluded that this observation was actually a target contaminant, which it was
not possible to establish in the measurements due to limitations in the focal plane detector.
We note further that the $2^+_2$ and $0^+_3$ states have been shown by the FMD
calculation \cite{Neff} to be nearly degenerate at the excitation energy
$E_{\rm x}\approx 11.8\sim 11.9$ MeV. Consequently, the probability
is high that the $2^+_2$ state is indeed the peak observed at
$E_{\rm x}\approx 11.46$ MeV in the 240 MeV \aap spectrum \cite{John03}.

\begin{table}\centering
\caption{Excitation energies and $E\lambda$ transition strengths of the IS states of $^{12}$C
under present study. Results of the AMD calculation \cite{Enyo07} are compared
with the available experimental data. Note that $M(E\lambda)$ is given in $e$~fm$^{\lambda+2}$
for 0$^+$ and 1$^-$ states; the experimental $B(E2;2^+_2\rightarrow 0^+_1)$ value has been
deduced from the $E2$ EWSR strength given in Ref.~\cite{John03} for the  $2^+$ peak
observed  at $E_{\rm x}\approx 11.46$ MeV.}
\vspace*{0.5cm}
\begin{tabular}{|c|c|c|c|c|c|c|} \hline
$J^{\pi}$ & $E_{\rm calc}$ & $E_{\rm exp}$ & Transitions & Calc. & Exp. & Ref. \\
  &(MeV) & (MeV)&  & (e$^2$fm$^{2\lambda}$) & (e$^2$fm$^{2\lambda}$)  & \\ \hline
2$^+_1$ & 4.5 & 4.44 &$B(E2;2^+_1\rightarrow 0^+_1)$ & 8.5 &$8.0\pm 0.8$ &  \cite{Ram01}\\
  &  &  &$B(E2;2^+_1\rightarrow 4^+_1)$ & 28.5  &  & \\ \hline
0$^+_2$ &8.1 & 7.65 &$M(E0;0^+_2\rightarrow 0^+_1)$ &6.7 &$5.4\pm 0.2$ & \cite{Stre70} \\
  &  &  &$B(E2;0^+_2\rightarrow 2^+_1)$ & 25.5  & $13.0\pm 2.0$ & \cite{Endt79} \\
  &  &  &$B(E2;0^+_2\rightarrow 2^+_2)$ & 511  &  & \\  \hline
3$^-_1$& 10.8 & 9.64 &$B(E3;3^-_1\rightarrow 0^+_1)$ &106 & $87.1\pm 1.3$ & \cite{Kib02} \\
  &  &   &$B(E3;3^-_1\rightarrow 2^+_2)$ &137&  &  \\ \hline
0$^+_3$&10.7&10.3 &$M(E0;0^+_3\rightarrow 0^+_1)$ &2.3 & & \\
&  &  &$B(E2;0^+_3\rightarrow 2^+_2)$ & 1553 & & \\  \hline
2$^+_2$ & 10.6 & $9\sim 11.5$   &$B(E2;2^+_2\rightarrow 0^+_1)$ & 0.4  &$0.5\pm 0.1$ & \cite{John03}  \\
  &   &   &$B(E2;2^+_2\rightarrow 0^+_2)$ &  102 & &  \\
  &   &   &$B(E2;2^+_2\rightarrow 4^+_1)$ &  13.5 & & \\
  &  &   & $B(E2;2^+_2\rightarrow 4^+_2)$ &  1071  &  & \\  \hline
1$^-_1$& 12.6 & 10.84 &$B(E3;1^-_1\rightarrow 2^+_2)$ &1679  & & \\
  &   &   & $M(E1;1^-_1\rightarrow 0^+_1)$ &2.56 &  & \\ \hline
\end{tabular}\label{t1} \\
\end{table}

To further investigate the excitation of the $2^+_2$ state in the \aap experiment,
we have used the AMD nuclear transition densities in our folding model analysis of
inelastic \aC scattering data measured with high precision at $E_\alpha=240$ MeV
\cite{John03} and 386 MeV \cite{Itoh04,Itoh08}. A generalized double-folding
method \cite{Kho00} was used to calculate the complex \aC potential as the following
Hartree-Fock-type matrix element of the complex CDM3Y6 interaction
\cite{Kho10,Kho97}.
\begin{equation}
 U_{A\to A^*}=\sum_{i\in \alpha;j\in A,j'\in A^*}[\langle ij'|v_{\rm D}|ij\rangle
 +\langle ij'|v_{\rm EX}|ji\rangle], \label{dfm}
\end{equation}
where $A$ and $A^*$ are states of $^{12}$C target in the entrance- and exit channels
of the \aC scattering, respectively. Thus, Eq.~(\ref{dfm}) gives the (diagonal) elastic
optical potential (OP) if $A^*=A$ and inelastic scattering FF if otherwise. The complex
density-dependent direct and exchange parts of the CDM3Y6 interaction $v_{\rm D(EX)}$
were taken the same as those parametrized recently \cite{Kho10} for the study
of \aPb scattering at 240 and 386 MeV. The accurate local density approximation
suggested in Refs.~\cite{Kho10,Kho01} has been used for the
exchange term in Eq.~(\ref{dfm}). All the DWBA and CC calculations have been performed
using the CC code ECIS97 written by Raynal \cite{Raynal}. The real and imaginary elastic
folded potential were scaled by the coefficients $N_{\rm R}$ and $N_{\rm I}$, respectively,
for the best optical model (OM) fit of the elastic scattering data:
$N_{\rm R}\approx 1.1, N_{\rm I}\approx 1.4$ and $N_{\rm R}\approx 1.3,
N_{\rm I}\approx 1.6$ for $E_\alpha= 240$ and 386 MeV, respectively. These same
$N_{\rm R(I)}$ factors were used to scale the real and imaginary inelastic folded FF
for the DWBA calculation, a standard method used so far in the folding + DWBA
analysis of inelastic \aA scattering \cite{John03,Kho00,Sat97}.
Since $N_{\rm R}$ and $N_{\rm I}$ are an approximate
way to take into account the higher-order (dynamic polarization) contributions
to the microscopic OP \cite{Kho00}, they must be readjusted again in the CC calculation
to account for those nonelastic channels that were not included into the CC scheme.
We then obtained  $N_{\rm R}\approx 1.1, N_{\rm I}\approx 1.2$ and $N_{\rm R}\approx 1.2,
N_{\rm I}\approx 1.3$ from the CC calculations for $E_\alpha= 240$ and 386 MeV,
respectively. These $N_{\rm R(I)}$ factors were also used to scale the complex inelastic
folded FF used in the CC calculation. The OM and CC descriptions of the elastic \ac
scattering at 240 MeV are shown in upper panel of Fig.~\ref{Elsig2}.
\begin{figure}
 \begin{center}
\vspace{-2cm}\hspace{0cm}
\includegraphics[angle=0,scale=0.70]{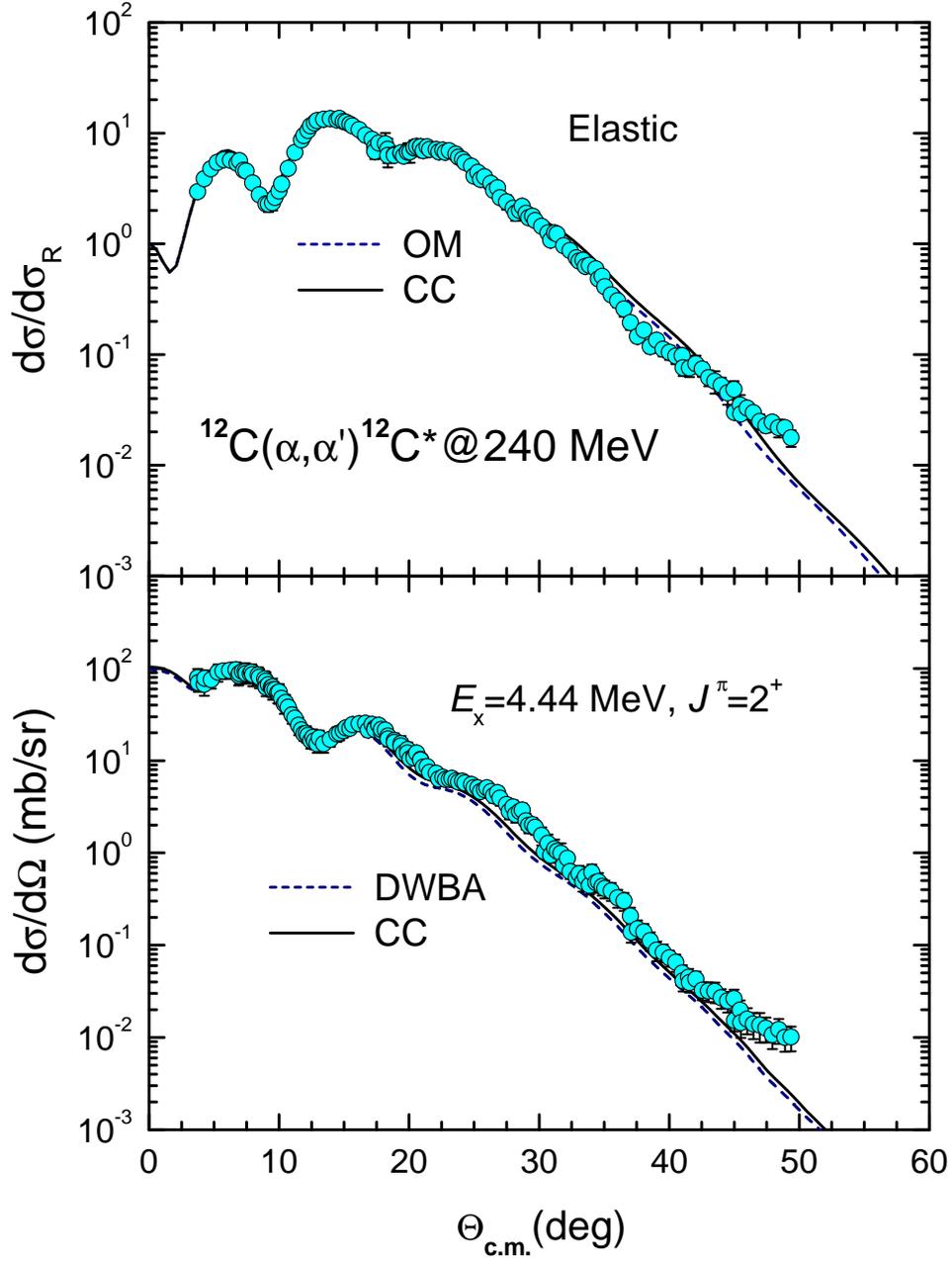}
  \vspace{-2cm}
\caption{Elastic and inelastic \ac scattering data at $E_\alpha=240$ MeV \cite{John03}
measured for the $2^+_1$ state at 4.44 MeV in comparison with the OM, DWBA and CC
results given by the complex double-folded OP and inelastic FF.}
 \label{Elsig2}  \end{center}
\end{figure}
Our OM calculation not only well describes the elastic data but also gives the total reaction
cross sections $\sigma_{\rm R}$ very close to the experimental values measured at the
nearby energies. Thus, the (complex) double-folded OP should be accurate enough for
the DWBA or CC analysis of inelastic \aC scattering. For the $2^+_1$ state, the electric
transition rate predicted by the AMD, $B(E2,0^+_1\to 2^+_1)\approx 42.5\ e^2$fm$^4$,
agrees perfectly with the measured value of $40\pm 4\ e^2$fm$^4$ \cite{Ram01}, and the
corresponding inelastic FF describes the measured \aap cross section quite satisfactory
in both the DWBA and CC calculations (see lower panel of Fig.~\ref{Elsig2}).  The calculated
\aap cross section for the $2^+_1$ state slightly underestimates the data at large angles and
this could well be due to a strong refractive effect that implies a weaker absorption
in the considered inelastic \aap channel \cite{Kho07}. While the $2^+_1$ state has been
observed in the spectra of all inelastic \aap experiments, the situation with the $2^+_2$ state
remains quite uncertain.
Given the possible peak of the $2^+_2$ state located at  $9.6\pm 0.1$ MeV as deduced by
Freer {\it et al.} from the \ppp spectrum \cite{Fre09}, in the same location as the first
$3^-$ state, it is highly suspected that the $2^+_2$ peak could have been merged with
the strong peak of the $3^-_1$ state and not observed in most of the measured
\aap spectra. In general, a $2^+$ state should have angular distribution oscillating
out-of-phase compared with that of the $3^-$ state and that effect could
well be revealed in the \aap angular distribution measured for the excitation energy
$E_{\rm x}\approx 9.6$ MeV if the $2^+$ cross section is strong enough.
\begin{figure}
 \begin{center}
\vspace*{-2cm}\hspace*{0cm}
\includegraphics[angle=0,scale=0.70]{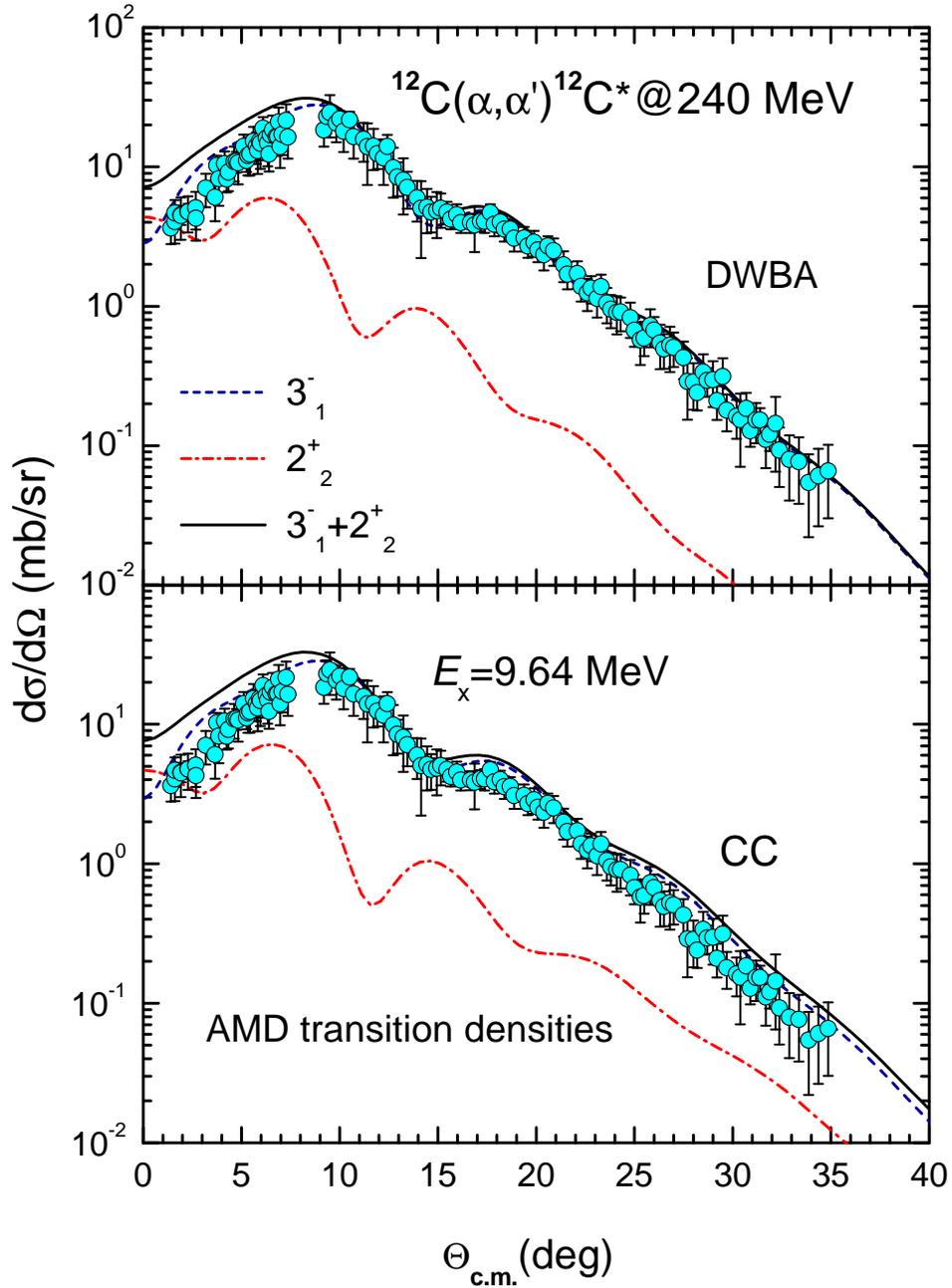}
  \vspace{-1.5cm}
\caption{Inelastic \ac scattering cross section measured for the $3^-_1$ state at
9.64 MeV \cite{John03} in comparison with the DWBA and CC results
given by the complex double-folded inelastic FF based on the AMD nuclear transition
densities for the $3^-_1$ and $2^+_2$ states.}
 \label{Insig3}  \end{center}
\end{figure}
To investigate this effect we have made the DWBA calculation of inelastic
\aC scattering at 240 MeV to the $3^-_1$ and $2^+_2$ states and the calculated cross sections
are compared with the data for the $3^-_1$ state in upper panel of Fig.~\ref{Insig3}. With the
AMD transition density giving the electric transition rate $B(E3)$ rather close to the
 measured value (see Table~\ref{t1}), the inelastic FF based on the AMD transition
 density describes the measured \aap cross section for the $3^-_1$ state quite well.
Compared to the $3^-_1$ cross section, the predicted inelastic scattering cross section for
the $2^+_2$ state is much weaker, with the ratio of  integrated \aC cross sections
$\sigma_{2^+_2}/\sigma_{3^-_1}\approx 12.8\%$. Such a strength ratio agrees reasonably
with the upper limit of about 15\% for the excitation strength of the $2^+_2$ state versus that of
the $3^-_1$ state deduced recently from the \ccp experiment at
$E_{\rm lab}=101.5$ MeV \cite{Bri10}.
\begin{figure}[bht]
 \begin{center}
\vspace*{-3cm}\hspace*{0cm}
\includegraphics[angle=0,scale=0.60]{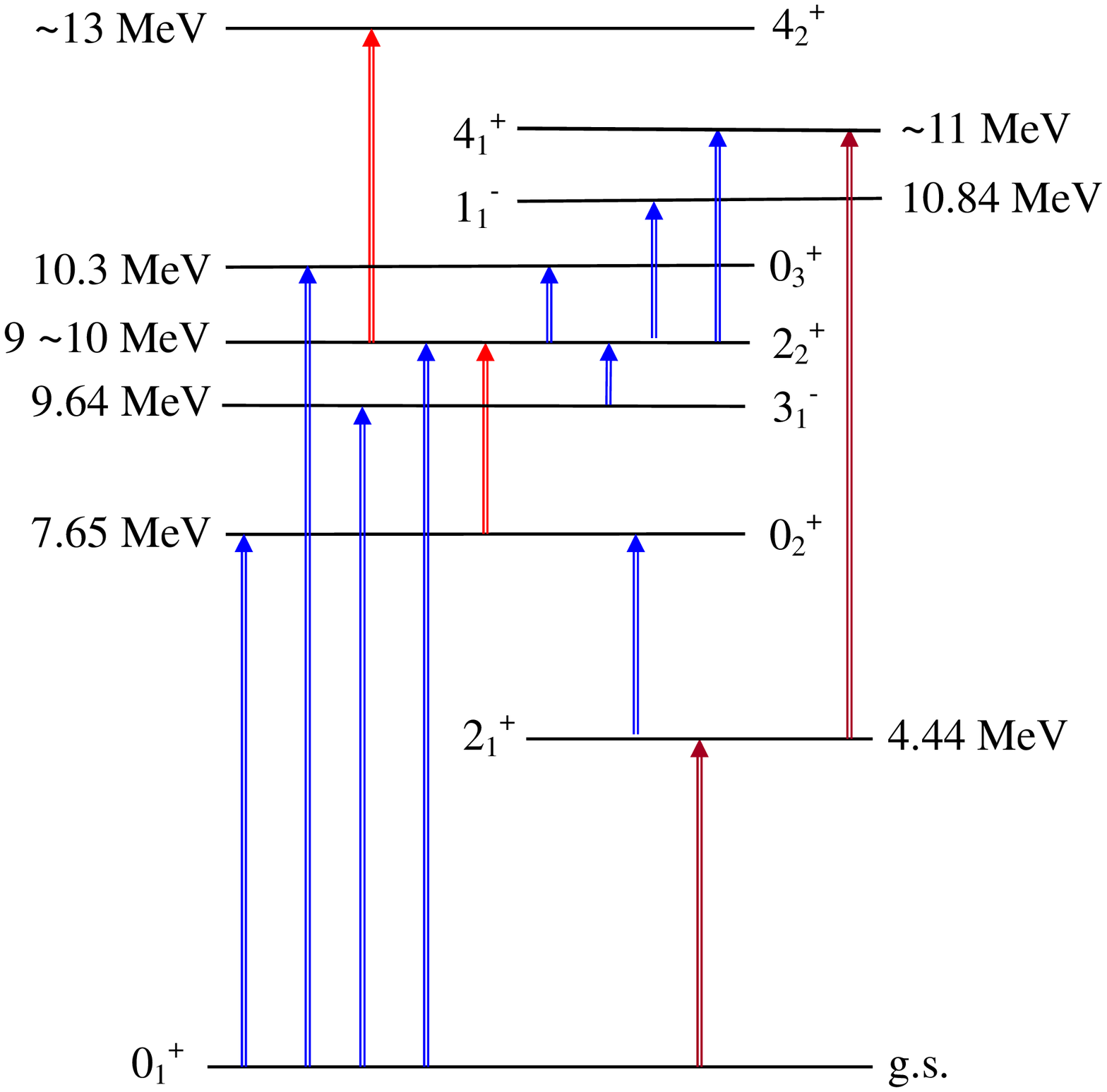}
  \vspace{-2cm}
\caption{Coupling scheme used in the CC equations for the elastic and inelastic
\ac scattering.}\label{cc}
   \end{center}
\end{figure}
Due to the \emph{reversed} oscillating pattern, the $2^+_2$ angular distribution is strongest
versus the $3^-_1$ one at the most forward angles. At angles
$\Theta_{\rm c.m.}\gtrsim 10^\circ$, the total $2^+_2+3^-_1$ cross section calculated
in the DWBA nearly coincides with the $3^-_1$ cross section. The \aap data points at forward
angles also indicate strongly that the data are indeed deduced for the $3^-_1$ cross section
and the contamination from  the $2^+_2$ state, if any, must be negligible. Therefore, in the
case of 240 MeV data the excitation strength of the $2^+_2$ state should be much smaller
than 12.8\% of the $3^-_1$ excitation strength if it is located at $E_{\rm x}\approx 9.6$ MeV.

It should be recalled that DWBA only treats the direct excitation and one needs to perform
the coupled-channel calculation in order to take into account contribution of the two-step
excitation of the $2^+_2$ state via the excited states of $^{12}$C lying around $9\sim 10$
MeV (see, e.g., very strong $E2$ transitions from the Hoyle state or  $0^+_3$ state
to the $2^+_2$ state in Table~\ref{t1}). For this purpose, we have computed the AMD
nuclear transition densities for all 13 transitions listed in Table~\ref{t1} and obtained
the corresponding inelastic scattering FF by the double-folding method (\ref{dfm})
for the CC calculation. The coupling scheme is illustrated in Fig.~\ref{cc} and the
CC results for inelastic scattering to the $3^-_1$ state are shown in lower panel
of Fig.~\ref{Insig3}. One can see that the contribution of the two-step excitation
of the $2^+_2$ state to the \aap cross section is rather small but not negligible.
It increases the $2^+_2$ cross section by at least 10\% and, hence, gives the ratio of
integrated cross sections $\sigma_{2^+_2}/\sigma_{3^-_1}\approx
14.8\%$. Because the $E3$ transition linking the $3^-_1$ and $2^+_2$ states is quite
strong (see Table~\ref{t1}), the CC effect also enhances slightly the $3^-_1$ cross
section. Nevertheless, it can be seen from Fig.~\ref{Insig3} that the direct (one-step)
$0^+_1\to 2^+_2$ excitation of the $2^+_2$ state is still dominant in the inelastic \aC
scattering at 240 MeV. Given a weak transition rate $B(E2,0^+_1\to 2^+_2)\approx
2\ e^2$fm$^4$ predicted by the AMD, the $2^+_2$ peak should be very difficult
to disentangle from the \aap spectrum if it stands just behind the strong  $3^-_1$ peak.
Moreover, the CC results and \aap data points at forward angles (lower panel
of Fig.~\ref{Insig3}) confirm consistently that the data points are indeed those for the
$3^-_1$ state and the mixture of the $2^+_2$ state should be negligible. In other words,
the contribution of the $2^+_2$ state to the inelastic \aap cross section at
$E_{\rm x}\approx 9.6$ MeV seems to be strongly suppressed in this case. To go down
in the beam energy might be a possibility to trace such a contribution because of stronger
CC effects. For example, our folding model analysis of the elastic and inelastic \aC data at
$E_\alpha=104$ MeV \cite{Hau69,Spe71} has shown that the CC effects substantially
increase the ratio $\sigma_{2^+_2}/\sigma_{3^-_1}$ (from 12.6\% in the  DWBA up
to about 27\% in the CC results). However, no angular distribution has been measured
for  the $3^-_1$ state at this energy and it is, therefore, difficult to make a similar discuss
about the $2^+_2$  state.
 \begin{figure}[bht]
 \begin{center}
\vspace*{3.5cm} \hspace*{-1cm}
\includegraphics[angle=0,scale=0.65]{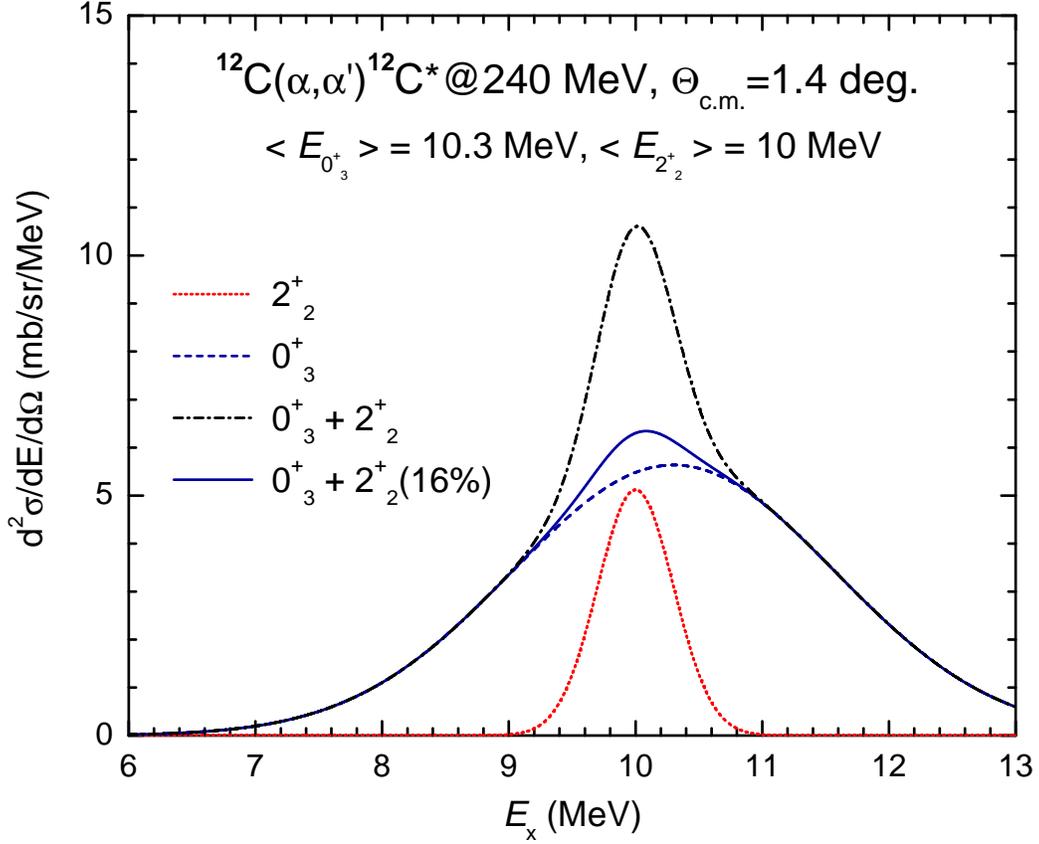}
  \vspace*{-6cm}
\caption{Decomposition of the 240 MeV \aap cross sections at $\Theta_{\rm c.m.}=1.4^\circ$,
predicted in the DWBA for the $0^+_3$ and $2^+_2$ states, into the Gaussians of 3 MeV and 0.6
MeV widths, respectively. See more details in text.}\label{mda240}
   \end{center}
\end{figure}

Situation becomes more complicated when we move up by about 500 keV in the excitation
energy to the $0^+_3$ peak at $E_{\rm x}\approx 10.3$ MeV. The predicted ratio
of integrated cross sections is $\sigma_{2^+_2}/\sigma_{0^+_3}\approx 81\%$ in
the DWBA calculation of \aap scattering at 240 MeV. Therefore, if the $2^+_2$ state
is located around 10 MeV, the \aap  cross section for the $2^+_2$ state should strongly
interfere with that of the $0^+_3$ state but such an effect has not been reported
experimentally \cite{John03}. To investigate this effect we have made an \emph{inverse}
multipole decomposition analysis by spreading the inelastic \aC scattering cross sections at
$\Theta_{\rm c.m.}=1.4^\circ$, predicted in the DWBA for the $0^+_3$ and $2^+_2$ states,
into the Gaussians of 3 MeV and 0.6 MeV widths, respectively, as deduced for the $0^+_3$
peak from the \aap spectrum at $E_\alpha=240$ MeV \cite{John03} and for the possible
$2^+_2$ peak from the \ppp spectrum  at $E_p=66$ MeV  \cite{Fre09}.
\begin{figure}
 \begin{center}\vspace*{-3cm}\hspace*{0cm}
\includegraphics[angle=0,scale=0.70]{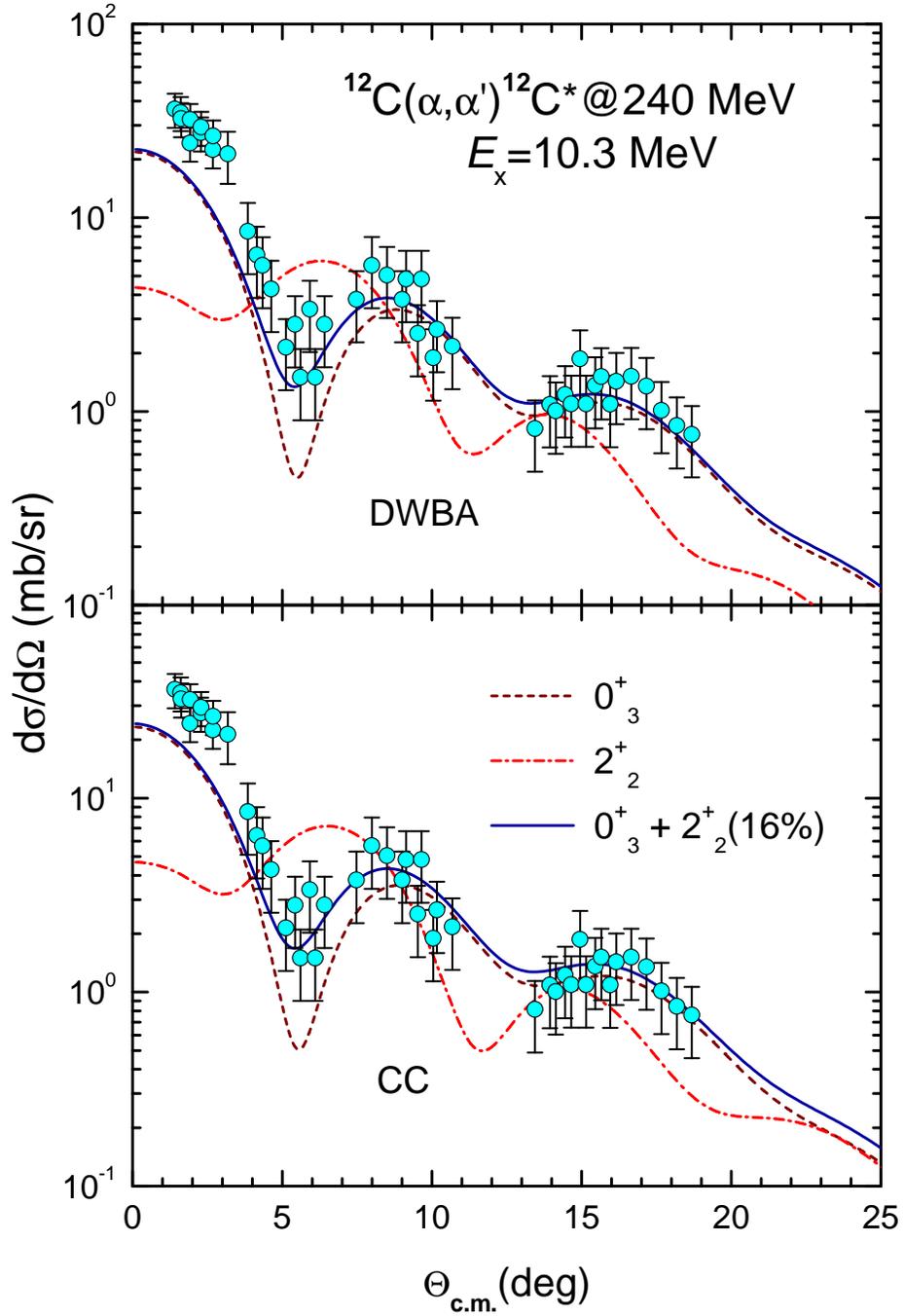}
\vspace*{-2cm}
\caption{Inelastic \ac scattering data measured for the $0^+_3$ state at
$E_{\rm x}\approx 10.3$ MeV \cite{John03} in comparison with the DWBA and CC cross
sections given by the complex double-folded inelastic FF based on the AMD nuclear
transition densities for $0^+_3$ and $2^+_2$ states. The total cross section (solid curve)
contains only 16\% of the  predicted $2^+_2$ cross section. }\label{Insig03}
   \end{center}
\end{figure}
The results of this decomposition analysis are plotted in Fig.~\ref{mda240} where the
area of each Gaussian has been normalized to the predicted DWBA cross section and
the centroids of the $0^+_3$ and $2^+_2$ peaks assumed to be around 10.3 and 10 MeV,
respectively. Given no strong interference between the $0^+_3$
and $2^+_2$ angular distributions observed in the 240 MeV experiment, we have tried
to trace the remnant of the $2^+_2$ state by reducing its strength in such a way that
the centroid of the sum of two Gaussians ($0^+_3 + 2^+_2$) remains within the experimental
value of $10.3\pm 0.3$ MeV as deduced from the 240 MeV data \cite{John03} for the
$0^+_3$ peak. Namely, by reducing the $2^+_2$ cross section to around 16\% of its predicted
strength at $\Theta_{\rm c.m.}=1.4^\circ$, we obtained the sum of the two ($0^+_3 + 2^+_2$)
Gaussians centered at $E_{\rm x}\approx 10.2$ MeV (see solid curve in Fig.~\ref{mda240}).
To further trace such a remnant of the $2^+_2$ state in the \aap cross section at 240 MeV
we have made the DWBA calculation using the inelastic FF's given by the full AMD transition
densities for the $0^+_3$ and $2^+_2$ states as well as the FF given by the AMD transition
density for the $2^+_2$ state scaled by a factor of 0.4 that corresponds to the 16\% reduction
of the $2^+_2$ cross section. From the comparison of these DWBA results with the
measured angular distribution for $0^+_3$ state in upper panel of Fig.~\ref{Insig03}
one can see that the AMD transition density for the $0^+_3$ state describes the data
quite reasonably and, hence, the monopole transition moment $M(E0)$ given by the
AMD (see Table~\ref{t1}) should be close to the realistic value.
Although rather small, the 16\% contribution of the $2^+_2$ cross section helps to
significantly improve the agreement with the data at the diffractive minimum around
$\Theta_{\rm c.m.}=6^\circ$.  The scaling of the AMD transition density for the
$0^+_1\to 2^+_2$ excitation by a factor of 0.4 shows the measure of suppression
of the $2^+_2$ state in this case. If we apply the same scaling to the mixture of the
$2^+_2$ state in the spectrum of the $3^-_1$ state then the ratio
$\sigma_{2^+_2}/\sigma_{3^-_1}\approx 2.4\%$ that is too small to be extracted
from the measured spectrum for the peak around 9.6 MeV. Independently, such a
conclusion can be well drawn from  the measured data shown in  Fig.~\ref{Insig3}.
Therefore, it is reasonable to assume in the CC calculation of the  \aap cross sections
for the $0^+_3$ and $2^+_2$ states at $E_{\rm x}\approx 10$ MeV the same scaling
for all inelastic FF corresponding to the transitions to and from the $2^+_2$ state
shown in Fig.~\ref{cc}. The CC results obtained with the scaled inelastic FF
are shown in lower panel of Fig.~\ref{Insig03}. We found that the coupling effects
enhance the $2^+_2$ cross section by about 50\%, with a slight change of
the $0^+_3$ cross section, and that leads to a much better agreement with the data
points over the whole angular region. The improved agreement with the data points
by the CC results indirectly indicate that the contribution from  the $2^+_2$ state
is not negligible as in the case of the $3^-_1$ state and it smoothens the measured
$0^+_3$ angular distribution at 240 MeV as shown in Fig.~\ref{Insig03}.
In other words, about 40\% of the predicted strength of the AMD wave function
$\Psi_{2^+_2}$ could be hidden in \aap spectrum measured at 240 MeV for the $0^+_3$
state. The AMD calculation has shown that the $0^+_3$ and $2^+_2$ states have quite
similar extended cluster structures (see Fig.~5 of Ref.~\cite{Enyo07}) and are almost
degenerate at $10.6\sim 10.7$ MeV. The most striking is a very strong ``interband"
transition $B(E2,0^+_3\to 2^+_2)\approx 1553\ e^2$fm$^4$, predicted by the AMD,
which helps to enhance the $2^+_2$ cross section by about 50\% in the CC calculation.
Thus, we have found that a \emph{ghost} of the $2^+_2$ state seems to be present
in the measured  $0^+_3$ angular distribution and following conclusions can be drawn
from our CC analysis of the 240 MeV \aap data:
 \begin{itemize}
 \item  If the $2^+_2$ state is located at the peak observed at $E_{\rm x}\approx 11.46$
 MeV then its width should be large enough to allow a tail of this peak to overlap with
 the broad $0^+_3$ peak. A direct CC analysis of the \aap cross section measured  for the
 energy bin centered at 11.46 MeV using the AMD wave function might solve this issue
 but no experimental angular distribution is available for that purpose.
 \item  If the $2^+_2$ state is located at $E_{\rm x}\approx 9\sim 10$ MeV, as predicted
 by some cluster calculations, then it should be hindered by the strong $3^-_1$ peak and
 only a weak fraction of its strength (about 16\%) is mixed with the broad $0^+_3$ peak.
\end{itemize}

\begin{figure}
 \begin{center}
\vspace*{-3cm} \hspace*{0cm}
\includegraphics[angle=0,scale=0.70]{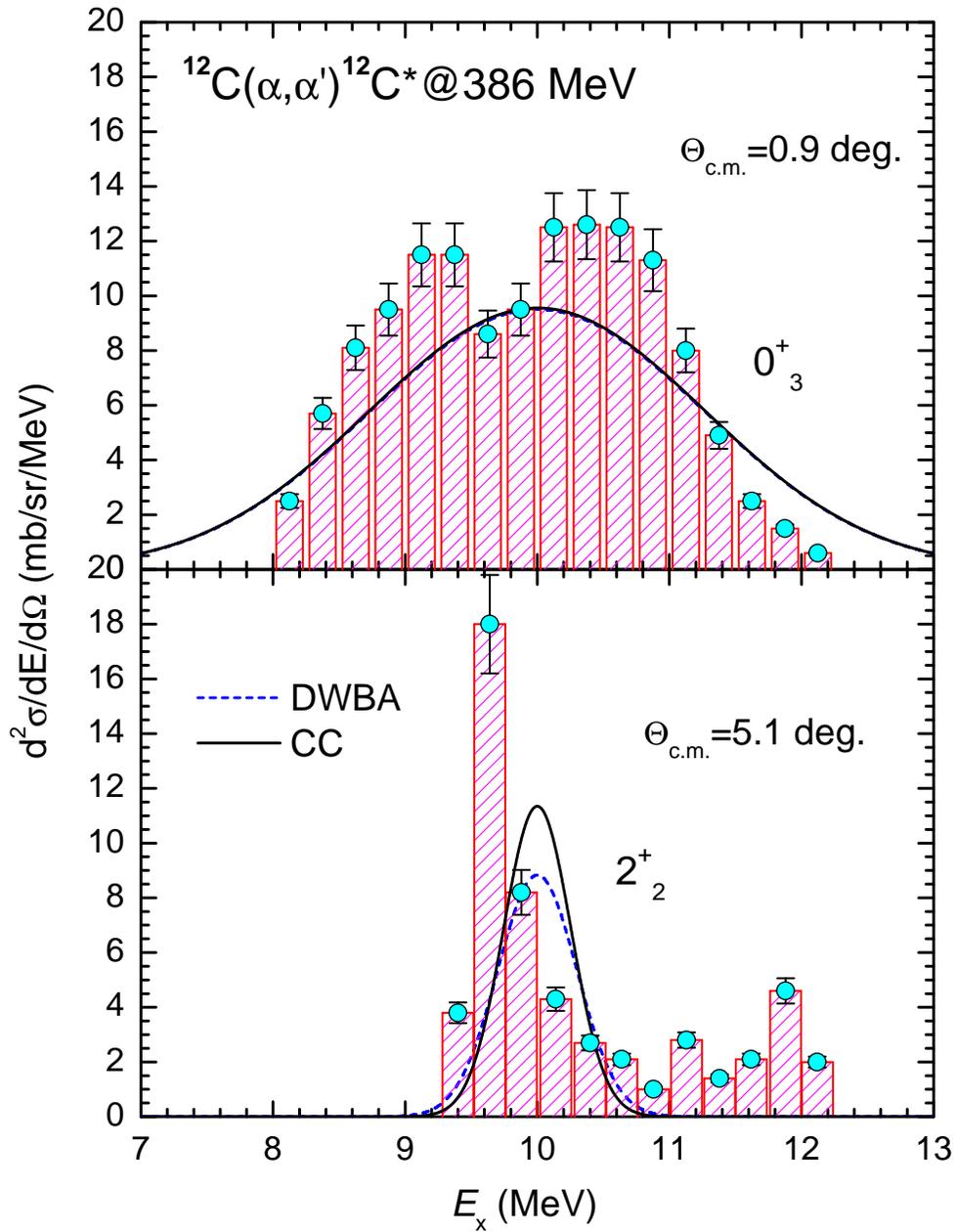}
  \vspace*{-1cm}
\caption{The same decomposition as in Fig.~\ref{mda240} but for the 386 MeV \aap cross
sections at $\Theta_{\rm c.m.}=0.9$ and $5.1^\circ$ predicted in the DWBA and CC
formalism in comparison with the measured data taken from Ref.~\cite{Itoh04}.
 See more details in text.}\label{mda386}
   \end{center}
\end{figure}

The measured  angular distribution has been subjected to a multipole decomposition
analysis to disentangle contribution of different multipolarities ($\lambda=0,1,2,3$) to the
excitation of $^{12}$C  in each energy bin, in the same way as done, e.g.,  in the inelastic
$\alpha$-scattering study of IS giant resonances in $^{208}$Pb \cite{Uch04}. It is,
therefore, possible to compare the predicted AMD transition strengths for the $0^+_3$ and
$2^+_2$  state with the experimental spectrum at some particular scattering angle.
For this purpose, we have done similar decomposition of the 386 MeV \aap cross sections
at $\Theta_{\rm c.m.}=0.9^\circ$ and $5.1^\circ$, predicted in the DWBA and CC
formalism for the $0^+_3$ and $2^+_2$ states, respectively, and the results are plotted
in Fig.~\ref{mda386} together with the corresponding double differential cross section
measured at $E_\alpha=386$ MeV \cite{Itoh04}. As can be seen in upper panel of
Fig.~\ref{mda386}, the AMD transition density for the $0^+_3$ state accounts fairly
well for the data points measured at the forward angle, with the integrated cross section
(over the excitation energy) $d\sigma/d\Omega\approx 30$ mb/sr compared to the experimental
value of around $33\pm 3$  mb/sr. Contrary to the situation for the $2^+_2$ state in
the 240 MeV case, the full transition strength predicted by the AMD still significantly
underestimates the observed strength (see lower panel of  Fig.~\ref{mda386}), with the DWBA
integrated cross section $d\sigma/d\Omega\approx 6.6$ mb/sr  at $\Theta_{\rm c.m.}=5.1^\circ$
compared to the experimental value of around $13\pm 2$  mb/sr.
 \begin{figure}[bht]
 \begin{center}
\vspace*{-2.5cm} \hspace*{-1cm}
\includegraphics[angle=0,scale=0.6]{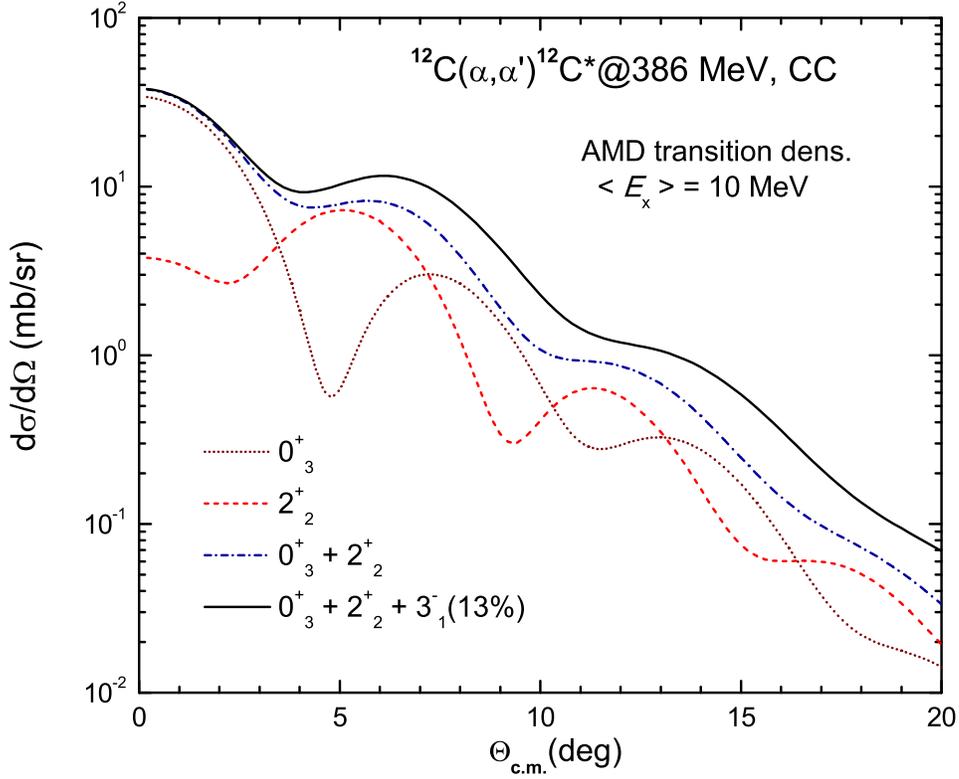}
  \vspace*{-1.5cm}
\caption{The CC results for the \aap cross section at 386 MeV obtained with the AMD
transition densities. The total cross section (solid curve) contains a 13\% contamination from
the $3^-_1$ state and agrees well with the experimental angular distribution
deduced for the excitation energy $\langle E_{\rm x}\rangle\approx 10$ MeV \cite{Itoh08}.
See more details in text.} \label{Insig386}   \end{center}
\end{figure}
It becomes obvious now that in the case of the $2^+_2$ state of $^{12}$C one has to deal
with about the same experimental difficulty as that in a study of isoscalar giant resonances, in
disentangling different IS excitation modes when their energies overlap. In this sense,
it is of interest to apply our AMD + folding approach to the inelastic \aC scattering data
measured at $E_\alpha=386$ MeV by Itoh {\it et al.} \cite{Itoh04,Itoh08}. We note that
these authors were able to measure the \aap energy spectrum without contamination
from the instrumental background by using the high-resolution magnetic spectrometer
Grand Raiden, and the \aap angular distribution has been deduced for each 250 keV
energy bin in the excitation energy range $3\lesssim E_{\rm x}\lesssim 20$ MeV
\cite{Itoh04,Itoh08}.

The CC calculation enhances the integrated cross section to $d\sigma/d\Omega\approx 7.3$ mb/sr
that is still well below the experimental value. Because the strongest peak in the experimental
spectrum  of the $2^+_2$ state is located at $E_{\rm x}\approx 9.6$ MeV, in about the same
position as that of  the $3^-_1$ state, it is not excluded that this experimental spectrum
has some contamination from the transition strength of the $3^-_1$ state \cite{Itoh10}.
Moreover, as shown in our recent folding model analysis of inelastic \aPb scattering to
the IS giant resonances in $^{208}$Pb, the IS transition strengths for a given
$2^\lambda$-pole excitation given by the MDA of the data measured at 240 MeV \cite{You04}
and 386 MeV \cite{Uch04} could be slightly different due to possible contribution from the
pickup/breakup reaction as well as different maximum $\lambda$ values taken into account
in the MDA. Keeping in mind possible uncertainty of the MDA, we have made an estimation
of the $3^-_1$ contamination in the experimental spectrum  measured at 386 MeV for the
$2^+_2$ state, based on the IS transition strengths predicted by the AMD \cite{Enyo07}.
To achieve a visually good agreement of our CC results with the preliminary data measured
at 386 MeV for the total $0^+_3 + 2^+_2$ cross section summed over the energy bins
around $E_{\rm x}=10$ MeV \cite{Itoh08}, we need to add to the predicted
$0^+_3 + 2^+_2$ cross section a significant contribution from the $3^-_1$ cross section.
These CC results are shown in Fig.~\ref{Insig386} and one can see that the
contamination by the $3^-_1$ cross section in the measured angular distribution
could be up to 13\% or more. Therefore, we conclude that at least 36\% of the predicted
strength of the wave function $\Psi_{3^-_1}$ could be hidden in the measured
386 MeV \aap cross section shown in Fig.~2 of Ref.~\cite{Itoh08}. Such a mixture of
the $3^-_1$ state could also affect the angular correlation function of the $\alpha$-decay
from the excited $^{12}$C$^*$ nucleus at $E_{\rm x}\approx 10$ MeV (see Fig.~3 in
Ref.~\cite{Itoh08}). Nevertheless, the fact that the full (direct and indirect) transition
strengths predicted for the $2^+_2$ state still underestimate the measured \aap spectrum
(see Fig.~\ref{mda386}) and angular distribution (see Fig.~\ref{Insig386}) indicates
that the authors of Refs.~\cite{Itoh04,Itoh08} were able to extract the full $E2$ transition
strength of the $2^+_2$ state from the 386 MeV \aap spectrum, even though the $2^+_2$
peak is located right behind the strong $3^-_1$ peak.

In conclusion, a detailed folding model analysis of the \aap data at 240 and
386 MeV in the DWBA and  CC formalism has been performed, using the nuclear
transition densities predicted by the AMD approach and a complex CDM3Y6 interaction.
From the structure point of view, given a very weak transition rate $B(E2;0^+_1\to 2^+_2)$
predicted by the AMD, the direct excitation of the $2^+_2$ state should be an unlikely event
in any reaction and that could be the reason why it was so difficult to identify the
$2^+_2$ state in the excitation energy- and/or $\alpha$-decay spectra of $^{12}$C.
Nevertheless, we have shown here some evidence for a \emph{ghost} of the $2^+_2$ state
in the 240 MeV \aap angular distribution measured at $E_{\rm x}\approx 10.3$ MeV,
which should be a tail of the $2^+_2$ peak located either at 11.46 MeV or right behind
the $3^-_1$ peak at 9.64 MeV. In addition to the weak transition $0^+_1\to 2^+_2$,
the strong $3^-_1$ peak was shown to be the main hindrance for the experimental
identification of the $2^+_2$ state.

The AMD transition densities account reasonably for the relative contributions
from the $2^+_2$ and $0^+_3$ states to the 386 MeV  \aap angular distribution measured
at $\langle E_{\rm x}\rangle\approx 10$ MeV.  We also found a contamination of about
13\% from the $3^-_1$ state in this angular distribution. Although the $2^+_2$ state
was found to be located near the strong $3^-_1$ peak, its full $E2$ strength has been carefully
deduced from  the 386 MeV \aap spectrum and these data \cite{Itoh04,Itoh08} remain so
far the only experimental evidence of the $2^+_2$ state at $E_{\rm x}\approx 10$ MeV.

Finally, going down in the beam energy might be an alternative to search for the $2^+_2$
peak in the \aap measurement because of very strong indirect transition $0^+_2\to 2^+_2$
that can be induced as a two-step excitation of the $2^+_2$ state in the CC scheme.
However, before discussing the indirect excitation of the $2^+_2$ state, we must check
which reaction channel is more likely for the Hoyle state: the direct $\alpha$-decay
or isoscalar $E2$ excitation. That should be an interesting perspective for a
future study of inelastic \aC reaction within the coupled reaction channel formalism.

Our study has been inspired by tireless experimental efforts by Martin Freer
to search for the $2^+_2$ state of $^{12}$C. We also thank Peter Schuck for his stimulating
and encouraging discussions. Communications with M. Itoh, T. Kawabata and X. Chen
on the measured \aap data are highly appreciated. The present research has been
supported, in part, by the National Foundation for Scientific and Technological
Development (NAFOSTED) under Project Nr. 103.04.07.09.

\end{document}